\documentclass[a4paper,11pt,epsf]{article}
\usepackage{amsmath}
\usepackage[psamsfonts]{amssymb}
\usepackage{rsfs}
\usepackage{bm}
\usepackage{eucal}
\begin{document} 

\hrule 
\leftline{}
\leftline{Chiba Univ. Preprint
          \hfill   \hbox{\bf CHIBA-EP-130}}
\leftline{\hfill   \hbox{hep-ph/0110013}}
\leftline{\hfill   \hbox{September 2001}}
\vskip 5pt
%\leftline{}
\hrule 
\vskip 0.5cm
\centerline{\Large\bf 
Vacuum condensate of mass dimension 2 } 
\vskip 0.5cm
\centerline{\Large\bf  
as the origin of mass gap and quark confinement$^*$}
\vskip 0.5cm

\vskip 1cm

\centerline{{\bf 
Kei-Ichi Kondo$^{\dagger}$,
}}  
\vskip 0.5cm
\begin{description}
\item[]{\it \centerline{ %$^1$ 
Department of Physics, Faculty of Science, 
Chiba University,  Chiba 263-8522, Japan}
}
\end{description}
%\vskip 0.5cm

{\bf Abstract:}
{\small 
This is a brief summary of recent works on the possibility of vacuum condensate of mass dimension 2 in Yang-Mills theory as the gluon sector of QCD.  In particular, I discuss the physical implications due to this condensate, especially, for the mass gap and quark confinement. This talk is mainly based on a recent paper \cite{Kondo01} and papers in preparation \cite{IKMS01}.
}

%\newpage
%\hrule  
%%%%% Table of Contents %%%%%
\pagenumbering{roman}
\tableofcontents
%%%%% Table of Contents %%%%%
%%%%%
%\baselineskip 23pt

\vskip 0.3cm  
\hrule  
\vskip 0.3cm  

$^*$  Talk presented at the workshop ``Structure and Reaction of Hadrons based on non-perturbative QCD" held at the Research Center of Nuclear Physics (RCNP), Osaka University, Japan, 23-24 July 2001.

\newpage

\pagenumbering{arabic}

\section{Introduction}
\setcounter{equation}{0}

\subsection{New vacuum condensates in QCD}

One of the well-known examples of vacuum condensate in QCD is
the quark condensate (of mass dimension 3) for light quarks:
\begin{equation}
  \langle 0| \bar{q}q |0 \rangle \not= 0 ,
\end{equation}
which implies the generation of the constituent quark mass associated with the spontaneous breaking of the chiral symmetry.
Another example is the gluon condensate (of mass dimension 4):
\begin{equation}
 \langle 0| \alpha_s (\mathscr{F}_{\mu\nu}^A)^2 |0 \rangle \not= 0 ,
\label{gluoncond4}
\end{equation}
which implies the existence of non-perturbative vacuum as suggested from the identity derived from the trace anomaly:
\begin{equation}
  \langle 0 | T_\mu^\mu | 0 \rangle 
= {\beta(\alpha_s) \over 4\alpha_s} 
\langle 0| (\mathscr{F}_{\mu\nu}^A)^2 |0 \rangle ,
\label{ta}
\end{equation}
where $T_\mu^\mu$ is the trace of the energy-momentum tensor $T_{\mu\nu}$ and $\beta(\alpha_s)$ is the renormalization group beta function with $\alpha_s:={g^2 \over 4\pi}$.  The RHS of (\ref{ta}) is negative for the asymptotically free theory characterized by the negative beta function $\beta(\alpha_s)<0$ when 
$\langle 0| (\mathscr{F}_{\mu\nu}^A)^2 |0 \rangle>0$.  Hence, the vacuum with a non-vanishing gluon condensate is more stable than the perturbative vacuum with $\langle 0 | T_\mu^\mu | 0 \rangle=0$.  
\par
Recently, two kinds of vacuum condensate of mass dimension 2 have been proposed independently.
The first one is the 
ghost condensation (in the Maximal Abelian gauge) \cite{Schaden99,KS00}
\begin{equation}
  \langle 0| i \bar{C}^a C^a |0 \rangle \not= 0 ,
\label{ghostcond}
\end{equation}
where the index $a$ runs over the off-diagonal components only.
The second one is the gluon condensation (in the Lorentz gauge) \cite{Boucaud00,GSZ01}
\begin{equation}
 \langle 0| \mathscr{A}_\mu^A \mathscr{A}_\mu^A |0 \rangle \not= 0 .
\label{gluoncond2}
\end{equation}

\subsection{Physical implications of vacuum condensate of mass dimension 2}

The non-vanishing vacuum condensates of mass dimension 2 
imply the existence of effective gluon mass and ghost mass.  This suggests the existence of mass gap in Yang-Mills theory.

Moreover, the vacuum condensates of mass dimension 2 imply the existence of the linear potential at {\it short} distance in the static quark potential:
\begin{equation}
  V(r) = - C_F {\alpha_s(r) \over r} + \sigma_s r ,
\end{equation}
where 
$C_F={N_c^2-1 \over 2N_c}$ and the string tension $\sigma_s$ is proportional to the vacuum condensate of mass dimension 2, e.g., for the Landau gauge in the Lorentz gauge fixing, 
\begin{equation}
  \sigma_s \cong g_R^2  \langle [\mathscr{A}_\mu \cdot \mathscr{A}_\mu]_R \rangle  .
\end{equation}
This should be compared with the conventional linear potential which is expected to hold in the long distance,
\begin{equation}
  V(r) =  \sigma_l r .
\end{equation}
This suggests the existence of a {\it short} confining string which is compatible with the asymptotic freedom.  
The available data of numerical simulations are not inconsistent with an observation,
\begin{equation}
  \sigma_s \cong \sigma_l .
\end{equation}
In the short distance, the coupling constant is small due to asymptotic freedom and hence it is possible to perform the reliable calculation.
Therefore, it is expected that quark confinement might be derived by extrapolating the short distance result to the long distance.  

Another advantage of this approach is that we can discuss the direct connection between the topologically non-trivial field configuration (monopole, instanton, meron, etc.) and the non-vanishing vacuum condensate. 
Even in the short distance (or high energy), a non-perturbative phenomenon is expected to occur (Note that the linear potential can not be derived in the perturbative expansion).

\subsection{Questions}
For the validity of the above scenario, we must answer the following questions.
\begin{enumerate}
\item
 The composite operators $\mathscr{A}_\mu^2$ and $\bar{C}C$ are neither gauge invariant nor BRST invariant.  Is there any physical meaning for such condensates?

\item
 Does the QCD vacuum really favor the vacuum condensate of mass dimension 2?

\item
 How does the existence of vacuum condensate affect the various physical quantities in QCD?  

\end{enumerate}

\section{Manifestly Lorentz covariant formulation of Yang-Mills theory (Review)}
\setcounter{equation}{0}
We analyze the Yang-Mills theory in the continuum formulation, not on the lattice.  Therefore, we begin with the manifestly Lorentz covariant formulation of Yang-Mills theory.
This material is well known and hence the details are omitted. 
 See the text book of quantum field theory.
The generating functional is given by
\begin{equation}
  Z[J,K] := \int \mathcal{D}\Phi e^{iS[\Phi]} 
\end{equation}
where $\Phi=\{ \mathscr{A}_\mu, \mathscr{C}, \bar{\mathscr{C}}, \mathscr{B} \}$ and
\begin{equation}
\begin{align}
  S[\Phi] &= S_{YM}^{tot} + S_J + S_K ,
  \\
  & S_{YM}^{tot} := - {1 \over 4} \mathscr{F}_{\mu\nu} \mathscr{F}^{\mu\nu} - i \bm{\delta} \left(\bar{\mathscr{C}}\left(\partial^\mu \mathscr{A}_\mu + {\alpha \over 2} \mathscr{B} \right) \right),
  \\
  & S_J := \int d^4 x \left[ J^\mu \cdot \mathscr{A}_\mu + J_C \cdot \mathscr{C} + J_{\bar{C}} \cdot \bar{\mathscr{C}} + J_B \cdot \mathscr{B} 
 \right] ,
  \\
  & S_K := \int d^4 x \left[ K^\mu \cdot \bm{\delta}\mathscr{A}_\mu + K_C \cdot \bm{\delta}\mathscr{C} \right] ,
\end{align}
\end{equation}
where $\bm{\delta}$ is the BRST transformation, $S_J$ is the source term for the elementary field and $S_K$ is the source term for the BRST transformed field.

\section{The Slavnov-Taylor identity}
\setcounter{equation}{0}

The BRST invariance of the theory is translated into a set of identities for Green's functions, the so-called the Slavnov-Taylor (ST) identities.  The ST identity is written in terms of the generationg functional of the connected Green functions and that of the 1PI Green functions.  
An important consequence of the ST identity is that the vacuum polarization tensor of the gluon is transverse.

\section{Choice of gauge fixing}
\setcounter{equation}{0}

By imposing a gauge fixing condition, the original local gauge symmetry $G_{local}$ is broken down to the residual local gauge symmetry $H_{local}$.
\begin{enumerate}
\item[1)] Lorentz gauge $\partial^\mu \mathscr{A}_\mu=0$:
$G_{local}$ is completely broken by the gauge fixing condition, $G_{local} \rightarrow H_{local}=\{ \emptyset \}$.  However, $G_{global}$ remains unbroken,
$G_{global} \rightarrow G_{global}$.

\item[2)] Maximal Abelian (MA) gauge:
$G_{local}$ is partially broken by the MA gauge fixing condition for the off-diagonal components, $G_{local} \rightarrow H_{local}$ where $H_{local}$ is the maximal torus subgroup of $G_{local}$.  At the same time,  $G_{global}$ is also broken explicitly,
$G_{global} \rightarrow H_{global}$.
See \cite{KondoI,KondoII} for details.

\end{enumerate}

\section{Mass gap in quantum Yang-Mills theory}
\setcounter{equation}{0}

The classical Yang-Mills theory is a massless and scale-invariant gauge theory.  In fact, the mass term ${1 \over 2}M^2 \mathscr{A}_\mu^2$ breaks the gauge invariance explicitly.

The quantum Yang-Mills theory is expected to have a nonvanishing effective mass.  This is due to the dimensional transmutation from the dimensionless parameters, $g$ (coupling constant) and $N_c$ (number of color), to the dimensionful scale $\Lambda$.  
Hence, non-zero mass $M$ proportional to the intrinsic scale $\Lambda$ is possible in quantum Yang-Mills theory.  
How to derive the mass theoretically?

1) The introduction of mass term by hand breaks the BRST invariance as well as the local gauge invariance explicitly.  

2) The introduction of Higgs scalar field can generate the mass through the non-vanishing vacuum expectation value of the Higgs field. However, this procedure alters the theory by adding extra scalar field.  The theory is no longer QCD.

3)  We examine the possibility of the dynamical mass generation. 

In view of this, we propose a linear combination of composite operators of mass dimension 2:
\begin{equation}
  \mathcal{O} = {1 \over \Omega} \int d^4 x \ {\rm tr} \left[ {1 \over 2} \mathscr{A}_\mu(x) \cdot \mathscr{A}^\mu(x) + \alpha i \bar{\mathscr{C}}(x) \cdot \mathscr{C}(x) \right] .
\label{O}
\end{equation}
where the trace is taken over the broken generators by the gauge fixing.
This operator is shown \cite{Kondo01} to be BRST and anti-BRST invariant, since
\begin{equation}
  \bm{\delta} \mathcal{O} = {1 \over \Omega}  \int d^4 x \partial^\mu {\rm tr}(\mathscr{A}_\mu \mathscr{C}) = 0 .
\end{equation}
In the Lorentz gauge, the Landau gauge $\alpha=0$ is a fixed point under the renormalization.  Therefore, the gluon composite operator is a meaningful quanity,
\begin{equation}
  \mathcal{O} = {1 \over \Omega} \int d^4 x \ {\rm tr} \left[ {1 \over 2} \mathscr{A}_\mu(x) \cdot \mathscr{A}^\mu(x)  \right] .
\end{equation}
This is also gauge invariant.
Therefore, the Landau gauge is a very economical gauge in the Lorentz gauge.
\par
On the other hand, this is not the case for the MA gauge, since $\alpha=0$ is not a fixed point in the MA gauge.  Therefore, in the MA gauge, we need to consider the full combination (\ref{O}), see \cite{IKMS01}.

\section{Operator product expansion}
\setcounter{equation}{0}

The operator product expansion for the gluon propagator (2-point function) 
$
  (G^{AB})(p) := \langle 0|T[\bar{\mathscr{C}}^A(p) \mathscr{C}^B(-p)]|0 \rangle
$
has been calculated:
\begin{equation}
\begin{align}
 (G^{AB})^{-1}(p) &= \delta^{AB}p^2  + \Pi^{AB}(p) ,
\\
  \Pi^{AB}(p) &= {N_c g^2 \over 4(N_c^2-1)} \left[
\langle (\mathscr{A}_\mu^A)^2  \rangle  
+   O\left( 1/p^2 \right) \right] .
\end{align}
\end{equation}
Owing to gluon condensation of mass dimension 2, we find that all the gluons acquire the nonvanishing effective mass as
\begin{equation}
  m_A^2 =  {N_c \over 4(N_c^2-1)}   
\langle g^2 \mathscr{A}_\mu^A  \mathscr{A}_\mu^A   \rangle  ,
\end{equation}
whereas all the ghost acquire the effective mass as
\begin{equation}
  m_C^2 = - {N_c \over 4(N_c^2-1)} 
\langle g^2 \mathscr{A}_\mu^A  \mathscr{A}_\mu^A   \rangle .
\end{equation}
Remarkably, the nonvanishing mass obtained for $\alpha=0$ in this way does not break the BRST symmetry, since the vacuum polarization tensor is transverse and the ST identity is always satisfied.  
However, the effective mass is gauge parameter dependent.%

Moreover, the OPE of the three-gluon vertex (3-point function) shows that the vacuum condensate of mass dimension 2 yields the UV corrections $\Lambda^2/Q^2$ in the QCD running coupling constant $\alpha_s(Q^2)$:
\begin{equation}
  \alpha_s(Q^2) = \alpha_s(Q^2)_{pert} \left[ 1+{g_R^2\langle \mathscr{A}_\mu^2 \rangle_R \over 4(N_c^2-1)}{9 \over Q^2} +O(\alpha) \right] .
\end{equation}
Such a correction for the coupling constant \cite{AZ98,BRMO98} leads to a piece of the linear potential 
$\sigma_s r$
at short distances $r \rightarrow 0$
in addition to the usual Coulomb-like potential $-C_F {\alpha_s(r) \over r}$ with
$C_F={N_c^2-1 \over 2N_c}$.  Here the string tension of the short-distance linear potential is proportional to the the gluon condensate of mass dimension 2:
\begin{equation}
  \sigma_s \cong g_R^2\langle \mathscr{A}_\mu^2 \rangle_R .
\end{equation}
The RHS of this equation is BRST and anti-BRST invariant (A special case $\alpha=0$ of the operator $\mathcal{O}$). So is the LHS of the string tension of the short string.
This argument suggests that the quark confinement is already intertwined in the UV region of QCD. 
Thus it is expected that the linear potential or the area law of the Wilson loop at short distances $r \ll 1$ could well be extrapolated to large distances $r \gg 1$ by including higher-order corrections up to a desired order.
\par
In the MA gauge, the OPE has been calculated to the lowest order, see \cite{Kondo01}.

%\section{Renormalization of composite operators}
%\setcounter{equation}{0}

\section{Effective action and potential of composite operator}
\setcounter{equation}{0}

In order to show that the QCD vacuum favors the vacuum condensation of mass dimension 2, we proceed to calculate the effective potential of the composite operator of mass dimension 2.
\par
In the course of calculation, we must take account of
\begin{enumerate}
\item
 renormalization of elementary fields:
\begin{equation}
\begin{align}
  & \mathscr{A}_\mu = Z_A^{1/2} \mathscr{A}_\mu^R, \quad
  \mathscr{C} = Z_C^{1/2} \mathscr{C}^R, \quad
  \bar{\mathscr{C}} = Z_C^{1/2} \bar{\mathscr{C}}^R, \quad
  \mathscr{B} = Z_A^{-1/2} \mathscr{B}^R, \quad
  \\
  & g = Z_A^{-3/2} Z_1 g^R, \quad
  \alpha = Z_A \alpha^R, 
\\
  & K_\mu = Z_C^{1/2}, \quad
  K_C = Z_A^{1/2} K_C^R .
\end{align}
\end{equation}
\item
 renormalization of composite operators:
For the composite operator, there is an additional divergence which cannot be removed by the renormalization constant $Z$ of the elementary field.  For example,
\begin{equation}
  [\mathscr{A}_\mu^2]_R = Z_{2} \mathscr{A}_\mu^2 .
\end{equation}
\item
 renormalization of vacuum energy divergence:
 If we include a source term 
$\omega \mathscr{O} = {1 \over 2} \omega \mathscr{A}_\mu^2$,
there appears a quadratic divergence in $\omega$ (This is a quartic divergence, since $\omega$ is of mass dimension 2). 
Therefore, we need a counterterm proportional to $\omega^2$, which spoils an energy interpretation.
(Moreover, the presence of ultraviolet renormalon implies Borel non-summability.)

\item
 operator mixing:
Given a source term $\omega_i \mathscr{O}_i$ for a composite operator $\mathscr{O}_i$, the renormalization of $\mathscr{O}_i$ is given by a linear combination of other composite operators,
$\Delta \mathscr{O}_i = Z_{ij} \mathscr{O}_i$.

\end{enumerate}

A correct way of introducing the composite operator $\mathscr{O}_i$ is as follows.
\begin{equation}
\begin{align}
  Z[J,K,\omega] &:= \int \mathcal{D}\Phi e^{iS[\Phi]} ,
\\
  S[\Phi] &= S_{YM}^{tot} + S_J + S_K +
  \int d^4x \left[ \omega_i Q_{ij} \mathscr{O}_j + {1 \over 2} \omega_i G_{ij} \omega_j \right] .
\end{align}
\end{equation}
where we have introduced the source $\omega_i$ for the composite operator $\mathscr{O}_i$ and the quadratic term in $\omega$ to absorb the vacuum energy divergence.
Inserting an unity,
\begin{equation}
  1 = N \int \mathcal{D}\rho_i \exp \left\{ i \int d^4x {-1 \over 2} (\rho - Q \mathscr{O} - G \omega)^\dagger G^{-1} (\rho - Q \mathscr{O} - G \omega) \right\} ,
\end{equation}
we arrive at
\begin{equation}
  Z[J,K,\omega,L] := \int \mathcal{D}\Phi \mathcal{D}\rho_i e^{i(S_{YM}^{tot} + S_J + S_K + S_\rho + S_\omega + S_L)} ,
\end{equation}
where
\begin{equation}
\begin{align}
  S_\rho &:= \int d^4x {-1 \over 2} (\rho - Q \mathscr{O})^\dagger G^{-1} (\rho - Q \mathscr{O}) ,
  \\
  S_\omega &:= \int d^4x \omega \rho ,
  \\
  S_L &:= \int d^4x L \bm{\delta}\rho .
\end{align}
\end{equation}
Thus, the term ${1 \over 2} \omega_i G_{ij} \omega_j$ quadratic in $\omega$ is cancelled to make the energy interpretation possible.  The term $\omega_i Q_{ij} \mathscr{O}_j$ was replaced by $\omega_i \rho_i$.
\par
The generating functional $W$ of the connected Green function is defined by
\begin{equation}
  W[J,K,\omega,L] = -i \ln Z[J,K,\omega,L] .
\end{equation}
By the Legendre transformation, the effective action, i.e., the generating functional of the one-particle irreducible (1PI) Green function is given by
\begin{equation}
 \Gamma[\underline{\Phi}, \underline{\rho}; K,L] := W[J,K,\omega,L] - \int d^4x [J \cdot \underline{\Phi} + \omega \underline{\rho}] .
\end{equation}
The effective potential for the composite operator is obtained by
\begin{equation}
  \Gamma[\underline{\Phi}(x)=0, \underline{\rho}(x)=\sigma, K=0, L=0] = - V(\sigma) \int d^4x .
\end{equation}

\section{Gluon condensation occurs!}
\setcounter{equation}{0}

For simplicity, we adopt the Landau gauge $\alpha=0$ in the Lorentz gauge fixing $\partial^\mu \mathscr{A}_\mu = 0$.
The renormalization factors for the elementary field are well known.
In the Landau gauge, there is no operator mixing for the composite operator
$\mathscr{O}={1 \over 2}\mathscr{A}_\mu \cdot \mathscr{A}^\mu$.
In the MS scheme of the dimensional regularization, the renormalization factor of the composite operator is calculated as
\begin{equation}
  Z_2 = 1 - {3 \over 2} {g^2N_c \over 16\pi^2} \epsilon^{-1} + \cdots .
\end{equation}
with $\epsilon=4-D$.
The counterterm of the vacuum energy divergence is calculated:
\begin{equation}
  \delta G = -3 {N_c^2-1 \over 16\pi^2} \epsilon^{-1} + \cdots .
\end{equation}

There is an arbitrary parameter $G$.  
We identify $G$ with a function of $g$ and we choose $G$ to be a unique meromorphic function of $g^2$ with a Laurent expansion around $g^2=0$:
\begin{equation}
  G(g^2) = {c_{-1} \over g^2} + c_0 + c_1 g^2 + \cdots .
\end{equation}
The coefficients $c_n$ can be determined so as to satisfy the renormalization group equation derived by requiring the multiplicative renormalizability.  
\par
The one-loop calculation of the effective potential $V(\sigma)$ shows that the potential has an absolute minimum at non-zero $\sigma$, $\sigma=\sigma_0$.
This implies that the gluon condensation of dimension 2 takes place and that the gluon acquires the effective gluon mass given by
\begin{equation}
  M^2 = \mu^2 \exp \left\{ - {16\pi^2 \over {13 \over 6}N_c g^2} \right\} .
\end{equation}
This form with an essential singularity at $g=0$ can not be obtained by the perturbative expansion up to a finite order, since all the coefficient of the Taylor expansion vanish at $g=0$.  
Two-loop calculation also supports the gluon condensation. 
See \cite{VKAV01} for two-loop results.

\section{Gauge parameter dependence and the Nielsen identity}
\setcounter{equation}{0}

By applying the Nielsen identity to our theory, we can conclude that
the non-vanishing vacuum energy $E_{vac}=-V(\sigma_0)$ due to vacuum condensate of mass dimension 2 is independent of the gauge fixing parameter $\alpha$, i.e.,
${\partial \over \partial \alpha}V(\sigma_0)=0$ when $V'(\sigma_0)=0$.
Therefore, we can estimate the gluon condensation of mass dimension 4 based on the identity (\ref{ta}):   
\begin{equation}
  \langle {\alpha_s \over \pi} \mathscr{F}^2 \rangle 
= -{32 \over 11} E_{vac} \cong 0.0031 \text{(GeV)$^4$} ,
\end{equation}
where we have used
$T_\mu^\mu = 4 E_{vac}= - 4 V(\sigma_0)$ and
$\beta(\alpha_s) = -{b_0 \over 2\pi} \alpha_s^2 + \cdots (b_0={11 \over 3}N_c)$.  Note that the RHS of the identity (\ref{ta}) is gauge invariant.  
The details will be presented in \cite{IKMS01}.

\section{Recent results of Lattice simulations}
\setcounter{equation}{0}

The vacuum condensate $\langle A_\mu^2 \rangle$ was searched by fitting the lattice data to the analytical expression obtained from the OPE up to 2-loop and 3-loop.  
The lattice data are obtained for the gluon propagator $G^{(2)}(Q)$ and the running coupling constant $\alpha_s(Q)$ defined by the (3-point) gluon vertex based on the MOM scheme.
The result of \cite{Boucaud00,Boucaudetal00,Boucaudetal01} is as follows. 
\begin{center}
 \begin{tabular}{l||c|c}
 $\sqrt{\langle A_\mu^2 \rangle}$ & 2-loop fit & 3-loop fit \\
 \hline
 $G^{(2)}(Q)$ & 1.67(17) GeV & 1.55(17) GeV \\
 $\alpha_s(Q)$ & 3.1(3) GeV & 1.9(2) GeV \\
 $\Lambda_{\bar{MS}}$ & 172(15) MeV & 233(28) MeV 
 \end{tabular}
\end{center}
Obviously, the 2-loop fit of the lattice data disagree with each other.
The 3-loop fit leads to the nearly equal value for the vacuum condensation.  Therefore, they draw a conclusion that the OPE at 3-loop order is reliable and the condensation is given by
\begin{equation}
  \sqrt{\langle A_\mu^2 \rangle} \cong 1.64(15) \text{GeV} .
\end{equation}

\section{Conclusion and discussion}
\setcounter{equation}{0}

In this talk, we have discussed the following issues.
\begin{enumerate}
\item
We have proposed a vacuum condensate of mass dimension 2 as the origin of mass gap and quark confinement in Yang-Mills theory.

\item
We have shown that there is a BRST and anti-BRST invariant combination
for manifestly Lorentz covariant gauges.  The Landau gauge $\alpha=0$ is a very special gauge where we have only to consider the gluon condensate in the Lorentz gauge.
This is not the case for the MA gauge.  This case is more involved.  

\item
Existence of such vacuum condensates has been suggested from the OPE calculation of gluon and ghost propagators (2-point functions) and vertex functions (3-point and higher point functions) in the Lorentz and the MA gauge \cite{Kondo01}.

\item
If the condensation in question really exists, gluons and ghosts acquire the non-zero effective mass.  Moreover, there is a linear potential at {\it short} distance, leading to a short confining string.

\item
It is possible to show that the Yang-Mills vacuum favours the gluon condensation based on the renormalizable effective potential of the composite operator ${1 \over 2}\mathscr{A}_\mu^2$.  The loop calculation has been done up to 2-loop order.

\end{enumerate}

In order to obtain the equivalent dual theory to the Yang-Mills theory with an insertion of the Wilson loop operator, we have introduced an antisymmetric (Abelian) tensor field as an auxiliary field of an gauge invariant antisymmetric tensor field which appears in the non-Abelian Stokes theorem for the Wilson loop, see \cite{Kondo00}.

At least in the short distance region, the obtained antisymmetric tensor gauge theory can be converted to 
\begin{enumerate}
\item
 the dual Ginzburg-Landau theory,

\item
 magnetic monopole-current theory,

\item
 the confining string theory
\end{enumerate}
by making use of the duality transformation, as suggested in \cite{KondoI} and demonstrated in \cite{Kondo00}.

The above results are expected to be extrapolated to the long distance by making use of the exact Wilsonian renormalization group method developed recently based on Polchinski's idea.

For phenomenological investigations on the gluon condensates of mass dimension 2, see \cite{CS86,GJJS86,CS84,SY99} and also \cite{AAEF97} for intimately related topics.%
\footnote{The author would like to thank D. Ebert and Martin Lavelle for bringing his attention to some of these references.}

%\newpage

\end{document}